\documentclass[acus]{JAC2000}

\usepackage{graphicx}

\begin{document}
\title{\flushright{THAP012}\\[15pt] \centering Online Model Server for the
Jefferson Lab accelerator\thanks{Work 
supported by the U.S. DOE contract No DE-AC05-84-ER40150}}

\author{Y. R. Roblin, T.L. Larrieu, Jefferson Lab, Newport News, VA23606, USA}

\maketitle

\begin{abstract}
A beam physics model server (Art++) has been developed for the Jefferson
Lab accelerator. This online model server is a redesign of the 
ARTEMIS model server \cite{artemis}. The need arose from an impedance mismatch
between the current requirements and ARTEMIS capabilities. The purpose
of the model server is to grant access to both static (machine lattice
parameters) and dynamic (actual machine settings) data using a single
programming interface. A set of useful optics calculations (R-matrix,
orbit fit, etc.) has also been implemented and can be invoked by
clients via the model interface. Clients may also register their own
dynamic models in the server. The server interacts with clients using
the CDEV protocol and data integrity is guaranteed by a relational 
database (Oracle8i) accessed through a persistence layer. 
By providing a centralized repository for both data and optics
calculations, the following benefits were achieved: optimal use of 
network consumption, software reuse, and ease of maintenance.
\end{abstract}

\section{Introduction}
The CEBAF accelerator is a 6 GeV, 5 pass, continuous beam electron 
recirculating linac \cite{CEBAF}. The two superconducting linacs are joined
head to tail by beam transport systems. The beam can be extracted during any 
of the five passes and delivered to three experimental halls. 
The purpose of Art++ is to provide an accurate and timely
representation of the machine for high level applications such as the
fast feedback system \cite{fastfeedback} or the automated beam steering system \cite{autosteer}. 
As such, Art++ serves as a central repository for beam physics
calculations (twiss functions, R-matrix, trajectories, etc.) and facilitates access to the set of parameters characterizing 
any beam element. By centralizing the source of information and beam
physics algorithms, we leverage beam physics expertise and our knowledge of machine parameters.

\section{Architecture}

\begin{figure}[ht]
\includegraphics*[width=80mm]{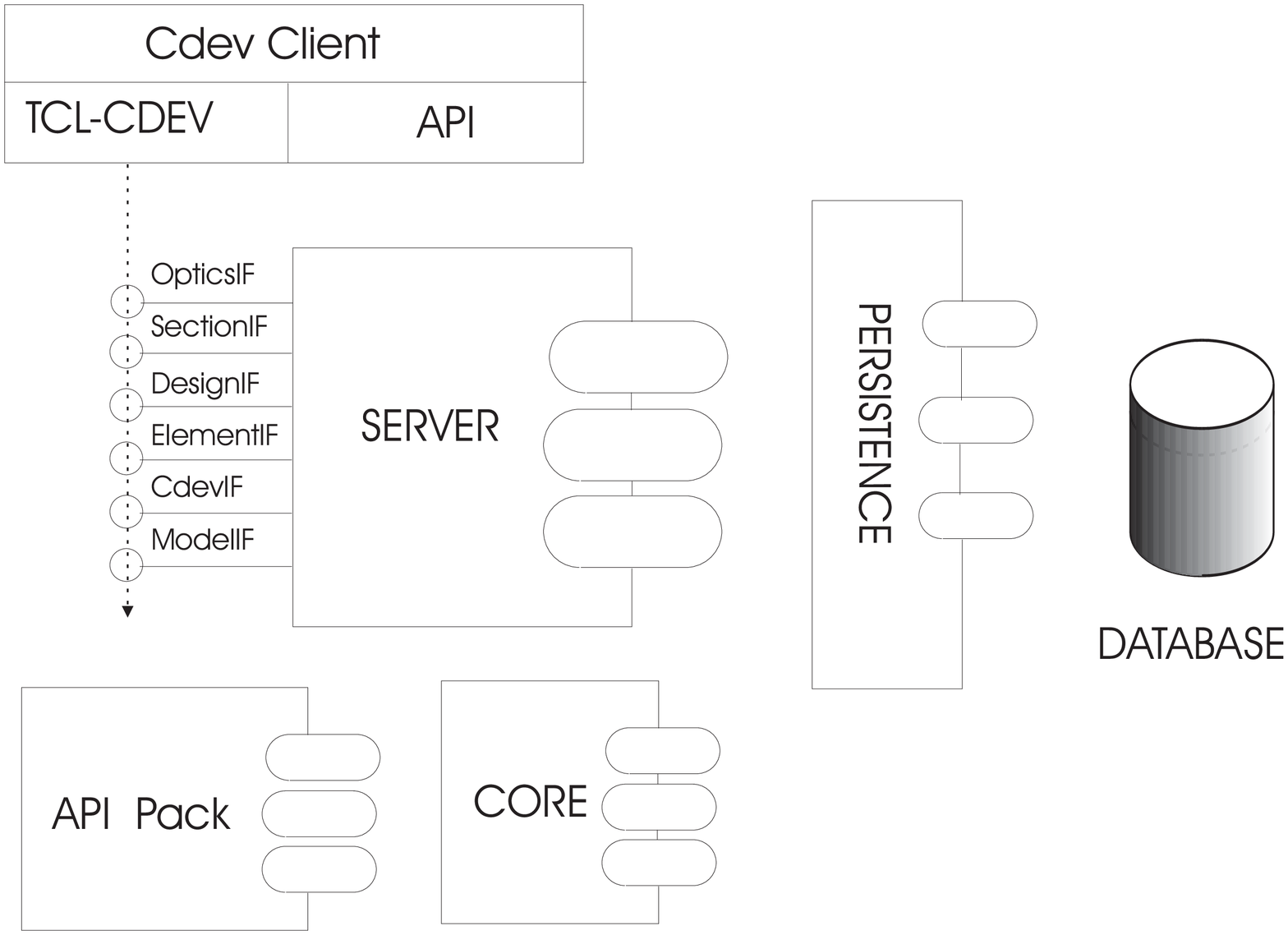}
\caption{Art++ physical view.}
\label{physview}
\end{figure}

Art++ is a distributed system, utilizing CDEV \cite{CDEV}
as its underlying protocol. The software architecture is
object-oriented, written in C++, and is organized in modules as shown in figure \ref{physview}. 
Each module is decoupled from the others and accessed only by means of
a well defined interface. The core module encompasses the business logic pertinent to the modeling of the accelerator.
The server module relies on CDEV protocol to interact with 
clients exposing a number of interfaces for the clients to act
upon. The API Pack module contains a number of parser classes to
interface with external data sources. The persistence layer provides
database functionalities to the core module. Currently, the supported
database mechanisms are relational database tables, implemented in Oracle, and flat files.

\subsection{Core package}
\begin{figure}[h]
\includegraphics*[width=80mm]{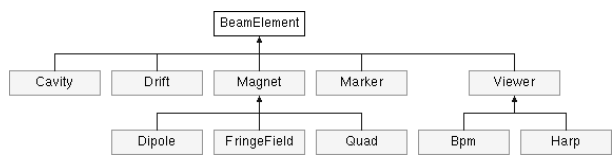}
\caption{Art++ class hierarchy.}
\label{hierarchy}
\end{figure}

The core package encompasses all the classes directly related to the
problem domain. The class hierarchy shown in  figure \ref{hierarchy}
 is closely matched to the real accelerator structure.
A model class holds a set of beam segments; each segment contains a
set of generic beam elements. More specialized versions of the beam
element class embody optical concepts such as quadrupoles, dipoles,
drift sections, etc.

\begin{center}
\begin{figure*}[t]
\center{\includegraphics*[width=120mm]{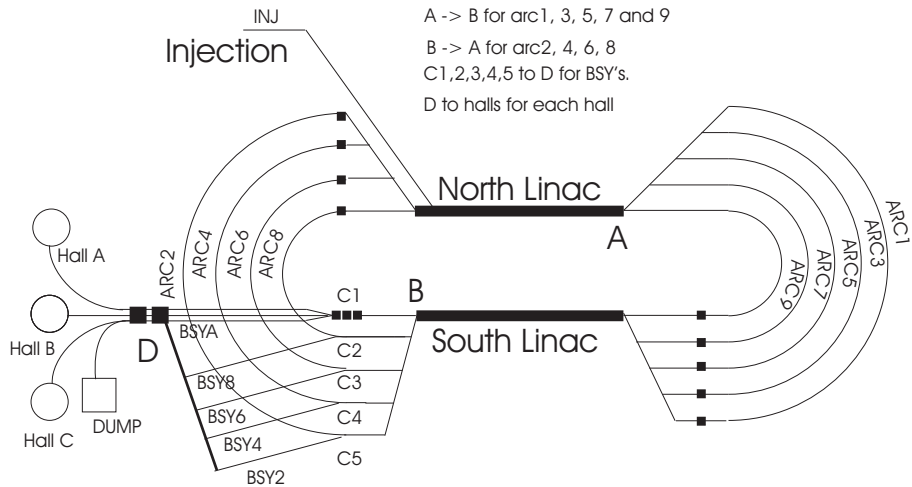}}
\caption{CEBAF accelerator lattice segmentation.}
\label{segmentation}
\end{figure*}
\end{center}

The segmentation of the beam line is shown in figure
\ref{segmentation}. Points of segmentation 
fall naturally at the beginning of each beam transport section delineated by the separation or recombination parts.

\subsection{API Pack package.}

Interfacing with external data sources is accomplished by the API Pack
package. It contains various parser classes to read machine
descriptions and currently supports only the Optim\cite{optim} and
twiss input files. Its framework can easily be extended to deal
with other formats. 
The accelerator lattice is specified by an xml \cite{XML} configuration
file which describes the beamline segmentation. Here is a typical configuration file 
for a short stretch of the machine, namely the injector and the first arc:
\small
\begin{verbatim}
<model>
<lattice name='CEBAF'>
<section name='INJ' energy='5.01268'>
<optfile>inj_nlp.opt</optfile>
<twissfile>inj_nlp.twiss</twissfile>
</section>
<section name='ARC1' energy='MMSARC1E'>
<optfile>arc1p.opt</optfile>
<twissfile>arc1p.twiss</twissfile>
</section>
</lattice>
<layout>
<pass number='1'>
<section>INJ </section>
<section>ARC1</section>
</pass>
</layout>
</model>
\end{verbatim}
\normalsize
Each segment is defined by its name and corresponding Optim
and twiss files that provide all the static
parameters needed by the model. We will refer to this set of data as
the ``design'' model in the rest of this paper.

\subsection{Server package}
The server package adapts the CDEV server interface to the internals of
Art++. It retrieves incoming messages, dispatching them to relevant
server methods and sending results back to the clients.

\subsection{TCL-CDEV interface}
Art++ can be queried via the TCL scripting language and TCL-CDEV
extension \cite{johannes}. The use of TCL scripting language
facilitates rapid prototyping \cite{rapid} of new applications. 
It can also be used for simple situations where performance is not a
stringent requirement. 

\subsection{API}
Alternatively, Art++ can be accessed via an API provided by
a C compatible shared library. We chose not to make it a 
C++ library with exported classes because of the need to support
legacy applications written in C. This API allows any user of the
control system to write thin-client applications making full use of the
model server. 
By utilizing CDEV as the underlying protocol, users can develop distributed 
applications with minimal effort.

\subsection{Persistence layer}
The role of the persistence layer \cite{amblerlayer} 
 is to encapsulate access to the
database, so that details about database table layout and corresponding SQL or
OCI (Oracle Call Interface) query mechanisms need not be coded into the
application. With the persistence layer in place the database can be
reorganized and the data access routines tuned without any modification to the
application itself.   The persistence layer uses a data dictionary to allow
mapping application objects to relational database tables and rows \cite{amblermapping}.
Although it might seem natural to use an object-oriented database to avoid such
object-to-relational mapping, we decided the effort of translation was
worthwhile because information stored in a relational database will be more
accessible and easier to maintain.  Information that needs to be updated often,
such as beamline survey data or magnet field properties, can be maintained by
simple scripts written by other staff.

\section{Beam physics computations}

Art++ provides availability to the following:
\begin{itemize}
\item Generation of first order matrices.
\item Calculation of twiss parameters.
\item Evaluation of transfer matrices at first order.
\item Correct treatment of the acceleration process in the linac including adiabatic damping and cavity focussing effects.
\item Determination of orbits.
\end{itemize}

All Art++ computations are carried out at first order. 
Transfer matrices are computed by using twiss parameters instead
of the alternative approach that calculates the product of all the 
matrices. This procedure results in an O(1) performance in transfer
matrix evaluation.

\section{Online realtime model}

To establish an accurate view of the machine, it is necessary to
track changes in the beamline in realtime. The dynamic parameters for
each dipole, quadrupole and cavity are monitored. Whenever one such
element changes, its local transfer matrix is recomputed and the
twiss parameters recalculated by forward
propagation from that point onward to the end of the segment. As in the case of the design model, 
the realtime model evaluates transfer matrices using twiss parameters. 

\section{High level applications}
Several high-level applications are using or being modified to use
Art++:
\begin{itemize}
\item Autosteer \cite{autosteer} is a multi-pass orbit correction
steering algorithm used at Jefferson Lab to steer the
machine. Autosteer determines the optimal corrector settings to
minimize beam displacement. It utilizes Art++ to
retrieve transfer matrices between beam position monitors and
correctors. 
\item Automatch \cite{automatch} performs beam envelope matching
between the transport segments. It requires knowledge of the
twiss parameters for the filling sections. Art++ computes these twiss
parameters in realtime for use by Automatch.
\item The fast feedback control program \cite{fastfeedback} corrects
the beam position and energy disturbances using a control loop based
on a linear quadratic gaussian controller/estimator. Art++ will be utilized
by the fast feedback system in its setup phase to retrieve twiss functions for various elements. 
\end{itemize}
 We plan to port all the high level applications to this new model
 server. Preliminary testing of the server was carried out concurently
 with the testing of Automatch. The first results show that the server
 behaves as expected. It exhibits a dramatic performance increase over
 Artemis, utilizing only a fraction of the model server CPU time (5 \% versus 80 \% for Artemis).

\end{document}